\documentclass[preprint,aps]{revtex4}
\begin{document}
\title{Of Connections and Fields-I}
\author{Joseph Samuel}
\affiliation{Harish Chandra Research Institute\\Jhusi, Allahabad,
India 211 019\\{\rm and}\\Raman Research Institute\\Bangalore, India 560 
080}
\date{\today}
\begin{abstract}

We describe some instances of the appearance of Chern's mathematical ideas
in physics.  By means of simple examples, we bring out the geometric and
topological ideas which have found application in describing the physical
world. These applications range from magnetic monopoles in electrodynamics
to instantons in quantum chromodynamics to the geometric phase of quantum
mechanics. The first part of this article is elementary and addressed
to a general reader. The second part is somewhat more demanding and
is addressed to advanced students of mathematics and physics.  
\end{abstract} 
\maketitle 

Take a loop of string and lay it out flat on a table so it doesn't cross
itself. Keeping your finger pointed along the string, move it all the way
around the loop at unit speed.  Note that the direction of your finger
describes a full circle.  Writing $\theta$ for the angle between your
finger and the $x$ axis, ($x$ and $y$ axes chosen conveniently along
perpendicular edges of the table) we find that the ``angular velocity''
$\kappa=d\theta/dt$ of the tangent vector (the unit tangent vector to the
string is ${\hat t}=(\cos \theta,\sin\theta$)) integrates to $2\pi$. We
now have the remarkable relation: \begin{equation} 1/2\pi \int_0^L \kappa
dt=1/2\pi \int_0^{2\pi} d\theta=1, \label{int*} \end{equation} where $L$
is the length of the string. What is so remarkable about this formula? If
one wiggles the string, the value of $\kappa$ at points of the string will
change. However, the integral (\ref{int*}) is unchanged. It is a
topological invariant, unaffected by small deformations. The angular
velocity $d\theta/dt$ of the tangent vector is just the curvature of the
string. Remember, we traverse the string at unit speed so, if $s$ is the
arc length parameter, $ds/dt=1$ and $\kappa=d\theta/ds=d\theta/dt$.
Physically, $\kappa$ is the transverse force necessary to keep on track a
bug of unit mass traversing the loop at unit speed. From its definition
$\kappa=d\theta/ds$, we see that $\kappa$ is a geometrical quantity,
described by a real number. However, its integral $1/2\pi\int\kappa ds$ is
an integer and a topological quantity, the number of times the tangent
vector ${\hat t}$ winds around the circle of directions in the plane. The
general flavour of the simple eq.  (\ref{int*}) is that the integral of a
certain geometric quantity over the loop is a topological quantity. This
is a simple example of a class of mathematical results discovered by the
mathematicians Chern, Weil and Pontryagin. These results are deep ones
involving global differential geometry and topology. They have had
applications in physics, in such diverse fields as elementary particle
physics and the quantum hall effect. The purpose of this article (and 
the next) is to
bring out the main impact of this body of mathematical work in the
physical sciences.  It is impractical within the space of these articles 
to
precisely define the mathematical objects that we deal with (manifolds,
connections, bundles). We are concerned less with completeness and rigour
than with enticing the reader into a beautiful subject. A full
appreciation of even this popular article would need some familiarity with
modern differential geometry and the standard model of elementary particle
physics. The hurried reader is advised to read this article
impressionistically and not worry too much about the technical details. A
less hurried reader may be motivated to try the suggested reading below to
pick up the necessary background to appreciate the mathematical ideas
described here and how they are used in physics. While reading this
article, you may want to keep pencil and paper handy so you can check some
of the formulae. Also required are string, a pair of scissors, glue (or
some less messy substitute like cellotape or a stapler) and a ball. Almost
any soft ball will do, like a tennis or basket ball.  If you are not the
sporting type, try and steal a ball from a child you know. Using your
hands to play with models will make the subject come alive as no
mathematical formulae can.

To continue with the elementary exposition, now consider a special case.  
Pull the loop of string taut in the shape of a triangle with vertices
$ABC$. The tangent vector ${\hat t}$ (and so $\theta$) remains constant
along the lines $AB$, $BC$ and $CA$ but swings abruptly through the
exterior angles $\alpha_1,\alpha_2,\alpha_3$ at the vertices $A$, $B$ and
$C$ respectively (Fig.1).  Adding up the total rotation of the tangent 
vector we
find that \begin{equation} 2\pi=(\pi-A)+(\pi-B)+(\pi-C) \label{triangle}
\end{equation} where we write $A,B,C$ for the interior angles at the
vertices $A,B,C$ respectively. From (\ref{triangle}) follows the equation
$A+B+C=\pi$ that you learned at your mother's knee: the sum of the
interior angles of a plane triangle is two right angles!  Elementary, yes,
but it has in it the germ of deeper ideas to come.
The reader is invited to puzzle over the following questions. What happens
if the string is laid on the table so that it crosses itself as for
example, a figure of eight?  What happens if the string is laid not on a
flat table but on the curved surface of a globe? How does one define 
the total rotation of the tangent vector? Is it still a topological 
quantity?  What happens if you suppose that the loop is infinitesimal so 
that the globe can be approximated by a plane? What happens if you 
expand the loop (say it is elastic) and shrink it back to a small loop
on the other side of the globe?
A little reflection will
show that these simple questions can be pursued in very interesting
directions. Let us now return to our main theme: the connection between
local and global quantities in mathematics and physics.

Consider a sphere $\cal S$ (you have one near you, the surface of the
ball) of radius $r$ embedded in three dimensional space. The curvature of
$\cal S$ (the Gaussian curvature which is also called the intrinsic
curvature) is given by $1/r^2$ and we have the easily checked result
\begin{equation} \frac{1}{4\pi} \int_{\cal S} \kappa dA=1
\label{gaussbonnet} \end{equation} where $dA=r^2\sin\theta d\theta d\phi$
in standard polar co-ordinates.  Now deform the ball by squeezing it. The
Gauss-Bonnet theorem states that (\ref{gaussbonnet}) still holds true. One
can understand this result as follows. At every point on the sphere ${\cal
S}$ with local co-ordinates $(x^1,x^2)$, there is a unit normal ${\hat
n}(x^1,x^2)$ which points out of the sphere. If $(x^1,x^2)$ are varied
over a small region of (signed) area $dA$, the unit normal varies over a 
signed area
$dA_{\hat n}$ on the sphere of directions. The Gaussian curvature is in
fact a measure of this variation and can be expressed as \begin{equation}
\kappa=\frac{dA_{\hat n}}{dA} \label{gausscurv} \end{equation} (The reader
who is not familiar with Gaussian curvature is invited to check for
herself (using (\ref{gausscurv})) that the Gaussian curvature is zero for
cylinders and planes\cite{foota}, positive for spheres and eggshells, and
negative for saddles and some parts of flower vases.) Thus, the Gaussian
curvature is a real number and a geometric quantity describing the local
curving of the sphere. However, its integral over the sphere is given by
\begin{equation} \frac{1}{4\pi}\int_{\cal S} \kappa dA=
\frac{1}{4\pi}\int_{\cal S} \frac{dA_{\hat
n}}{dA}dA=\frac{1}{4\pi}\int_{\cal S} dA_{\hat n}=1
\label{gaussbonnetproof} \end{equation} since the unit normal ``winds once
around'' the sphere of normals when the sphere is traversed once. Just as
in the case of the loop of string, we find that although the integrand is
local and geometrical, the integral is a topological quantity which being
an integer, does not change under small variations.  While deforming the
ball one creates non- uniformities in the curvature. However, every
decrease in the curvature is compensated by an increase somewhere else so
that the integral remains unchanged.
 
The Chern Weil theorem is a general mathematical result of a similar
nature dealing with connections on fibre bundles. A fibre bundle is a
space which is locally a product of two spaces. A simple example of a
fibre bundle is a M\"obius strip. To make your very own bundles, you can
use a xerox copy of Figure 2. Cut along the dark black lines to get four
strips. Taking a strip at a time, hold the two ends of each strip against
the light and align the paper so that the square, triangle and circle
overlay the square, triangle and circle at the other end. Make sure that
the triangle is properly oriented. For the second strip (B) you will have
to turn one of the ends around once to do this. Glue the ends in place.
You are now the proud owner of four fibre bundles, labelled A,B,C and D!
Take good care of them; they may be worth a bundle someday. Let us compare
the first two, A and B a cylinder and M\"obius strip respectively.  
Locally, the M\"obius strip B is indistinguishable from the cylindrical
strip A (one without the half twist), but it is globally different.  A bug
walking along a small part of the dotted line cannot tell the difference,
but we know that the two are quite different spaces. For example, the
M\"obius strip has only one side (try colouring it) as opposed to two
sides for a cylindrical strip. If you cut A and B all the way around along
the dotted line, you will find that A falls into two components, but B
does not! (What happens if you repeat this operation on B?) The M\"obius
strip is a twisted product of a circle $S^1$ and a line.  The circle is
called the base of the fibre bundle and the line the fibre. More
generally, a fibre bundle consists of $({\cal E}, {\cal B}, {\cal
F},\Pi)$, where ${\cal E}$ is the total space and $\Pi$ is a projection
from ${\cal E}$ to the base space ${\cal B}$ and ${\cal F}$ is the fibre.  
We require that any point $b\in {\cal B}$ has a neighbourhood ${\cal U}_b$
so that ${\Pi}^{-1}({\cal U}_b)$ looks like ${\cal U}_b\times {\cal F}$.
This is the requirement that the space is locally a product space.  It may
happen that globally ${\cal E}={\cal B}\times {\cal F}$, in which case, we
say that the bundle is trivial. More generally, the bundle is only locally
a product not globally.  The mathematical language of fibre bundles turns
out to be just right for describing a class of physical theories called
gauge theories.  Gauge theories have turned out to be extremely successful
in describing the interactions between elementary particles.  The most
familiar example of such a theory is electromagnetic theory, that you may
have studied. Mathematically, gauge theories (more precisely classical
gauge theories) are connections on fibre bundles. A connection gives a
rule for ``parallel transport'' or moving between fibres.  See the
accompanying article in this issue by Siddhartha Gadgil for more on this.
You can visualise a connection by using the paper strip models C and D,
which are fibre bundles with a connection on it. The fibres are shown as
short grey lines across the strip. The rule for horizontal motion in each
of these bundles is: move parallel to the dark thin lines. With the bundle
C you find that going all the way around the strip brings you back to the
same point on the fibre: the connection is integrable. With the bundle D
you come back to a {\it different} point on the same fibre. This
connection is not integrable. Integrable connections are flat and have
vanishing curvature. Locally integrable connections have vanishing
curvature, but there may be global obstructions to integrability as in the
case of the M\"obius strip. Can you put an integrable connection on a
M\"obius strip? Try it out and see.

We now turn to a simple example of Chern's ideas in electromagnetic
theory:
the quantisation condition for the charge of magnetic monopoles. A 
magnetic 
monopole is an object from which magnetic field lines emerge (just 
as electric field lines emerge from an electrically charged particle such 
as an electron). 
Magnetic monopoles have never been seen experimentally, but have
fascinated physicists over many generations. J.J. Thomson, P.A.M. Dirac,
M.N. Saha have studied monopoles in years past and more recently,
G. `tHooft and A.M. Polyakov have made seminal contributions to the 
subject. 
Dirac showed that if even a single monopole exists, the electric charges
of all partices  must be multiples of a certain basic unit of
charge  $q=n(\frac{\hbar}{(2g)}$ or 
\begin{equation}
qg=n\hbar/(2)
\label{quantisation}
\end{equation}
where $g$ is the magnetic charge of the monopole and $\hbar$ is Planck's
constant.
Even though monopoles have not been seen, {\it quantisation of electric
charge is a fact of Nature}. This lends support to the idea that
magnetic monopoles {\it may} exist. In fact, the standard model
of the weak nuclear interactions 
predicts monopoles of the kind studied by `tHooft and Polyakov.

In the modern gauge theoretic approach towards electromagnetic theory,
one views the magnetic fields (and electric fields) as analogous to
curvature. The integral of the magnetic field over a closed surface ${\cal 
S}$
measures the total magnetic charge contained within that surface.
\begin{equation}
(2q/\hbar)\int_{\cal S} {\vec B}.d{\vec S}=n
\label{flux}
\end{equation}
Thus we see by analogy that the general flavour of Dirac's quantisation 
condition (\ref{quantisation}) is the same as the Chern-Weil result:
the integral of the curvature over a closed surface is an integer!

{\bf The Geometric Phase}: One physical context where connections appear
naturally in physics is the geometric phase. This is a vast topic,
interesting in its own right. Let us briefly recall the main ideas in the
context of the adiabatic theorem of quantum mechanics. A quantum system in
a slowly changing environment displays a curious history dependent
geometric effect: when the enviroment returns to its original state, the
system also does, but for an additional phase. The phase is a complex
number of modulus one and experimentally observable by quantum
interference. This phenomenon is of geometric origin and is fundamentally
due to the curvature of the ray space of quantum mechanics. (Ray space is
the Hilbert space modulo multiplication by non-zero complex numbers.) The
geometric phase provides us with a wealth of examples of globally
non-trivial bundles, monopoles, instantons and more.

A quantum state 
$|\psi>$ is an element of a complex Hilbert space ${\cal H}$,
(let's say finite dimensional ${\cal H}={\rm I \!\!\! \rm C}^N$ to keep life simple)
with the inner product $<\phi|\psi>=\Sigma_{i=1}^N {\bar 
\phi}_i\psi_i$. The time evolution of the system is determined by the
Schr{\"o}dinger equation 
\begin{equation}
i{\frac{d}{dt}}|\psi>={\widehat{H}} |\psi> 
\label{schrodinger}
\end{equation}
where ${\widehat{H}}$ is an $N\times N$ Hermitian matrix. Let us suppose 
that the Hamiltonian ${\widehat{H}}$ depends on a set of parameters 
$\{x^1,x^2...x^m\}$ (which we write collectively as $x$, not
to be confused with the ordinary spatial co-ordinate) describing
the influence of the environment on the system. 
For $x$ 
fixed, the 
equation
\begin{equation}
{\widehat{H}}(x)|\psi(x)>=E(x) |\psi(x)>
\label{eigen}
\end{equation}
gives us the eigen states of ${\widehat{H}}(x)$. From the 
Schr\"odinger equation, the eigenstates of 
${\widehat{H}}$ evolve by a pure phase ${\exp-iE(x) t}|\psi(x)>$. This is 
the 
dynamical phase, which is easily accounted for. Consider a region of
parameter space where the eigenvalues of ${\widehat H}$ are well
separated from each other. Then $\tau=\hbar/(E-E')$  is small for 
any pair $E,E'$ of eigenvalues. If the parameter $x^i$ is slowly
varied (over a time large compared to $\tau$), a system which
is initially in an eigenstate 
remains in an eigenstate of the instantaneous Hamiltonian. 
Further, since the energy levels do not cross, one can unambiguously
keep track of the energy levels. For example, the eigenspace corresponding
to the lowest energy level is a well defined notion.
If the parameters $x$ are varied in
a cyclic fashion, one returns to the original ray. The eigenvalue
equation(\ref{eigen}) determines a ray in ${\cal H}$ for every $x$. 
This is a fibre bundle over the parameter space. This fibre bundle
has a natural connection which can be deduced from the 
Schr\"odinger equation: move orthogonal 
to the fibres in the 
Hilbert space inner product. 
\begin{equation}
<\psi|{\frac{d}{dx}}|\psi>=0
\label{transport}
\end{equation}
The curvature of this connection
is Berry's phase. Berry's phase has been seen experimentally in 
numerous situations. The simplest example is a spin half system
which is described in the Box. The connection occurring here is the
same as that of the magnetic monopole.

We have come a long way from the loop of string
we started with! The elementary results we used to motivate this
exposition are only the tip of an iceberg. The basic idea is
capable of considerable generalisation. Many of these ideas have
application in modern physics.
Such results find application in understanding gauge theories in
many contexts, from Chern-Simons gauge theories of the quantum Hall effect 
to the non-Abelian gauge theories of the standard model of elementary
particle physics. 

While one may object that magnetic monopoles
are ``science fiction'' from the experimental viewpoint, 
the theoretical idea is very powerful and useful. The monopole 
configurations appear naturally in the geometric phase of a two state
quantum system. With the right mathematical mappings,
magnetic monopoles can be usefully employed to understand
the elastic properties of DNA! (see suggested reading below).
Even if magnetic monopoles are never directly seen in the laboratory,
the theoretical ideas they bring with them are very deep and capable
of application. Physics is a deeply 
interconnected subject and we can transport  
theoretical ideas from one context to another.

It is curious that the deep mathematical ideas of Chern-Weil theory
which were discovered with a purely mathematical motivation actually
turn out to be useful in physics \cite{wigner}. Nature makes liberal use 
of geometric
and topological objects. All the known interactions of Nature use the
idea of connections on fibre bundles (gauge theories to physicists). 
This influx of mathematical ideas into physics is a good example of
the symbiotic relation between the two disciplines. 
Mathematical ideas often find applications in physics and 
these applications provide a further stimulus for development of
the subject.

The reader may wonder why there is so much mathematics in physics.
The physics that we were taught at high school was of a more
descriptive variety, with boring definitions to remember and the shape of
some round bottomed flask used in some long forgotten experiment! The
present state of the subject is considerably different and theoretical
physicists need to have a fairly good base in mathematics to even get
started. This is due to the natural evolution of the subject and the fact
that both theory and experiments are getting more and more sophisticated.  
Throughout history, each generation of physicists has complained that the
next generation is far too mathematical. (In the 1930s, physicists used to
complain of the `gruppenpest'', the invasion of physics by group theory.)
Physicists freely borrow state of the art mathematical language to
formulate their theories.  However, it is important to remember that
theoretical physics is not mathematics. It uses mathematics to describe,
understand and predict the behaviour of the physical world. The use of
mathematics is not an end in itself, but a means to an end: a better
understanding of Nature.

{\it Acknowledgements:} It is a pleasure to thank Indranil Biswas, Rukmini
Dey, Rajesh Gopakumar, H.S. Mani, Sinha, Supurna Sinha, B. Sury and K.P.
Yogendra for reading through this article and helping me to improve it.  
\newpage
                                                                                                                                                                                                                                                                                                                                                                                        
{\center{\large Box: Monopoles, M\"obius strips and Berry's Phase}}\\
Let us consider a very simple example in which a globally non-trivial
connection (gauge field) appears: a spin $1/2$ system in an external 
magnetic field. We regard the three components of the magnetic field
as parameters $x^i,i=1,2,3,$ which can be varied.
Throughout this box, the parameters $x^i$ represent co-ordinates on 
the base space ${\cal B}$ which is the parameter space 
describing the influence of the environment, and not 
space-time as was the case in the previous examples.
We write the Hamiltonian, a $2\times2$, Hermitian matrix  
 as $H=x^i\sigma_i$, where $\sigma_i$ are the three Pauli matrices. These
matrices are traceless and Hermitian and satisfy
$\sigma_i\sigma_j+\sigma_j\sigma_i=2 \delta_{ij}$, the Clifford algebra.
Since the eigenspaces of $H$ are the same for $x^i$ and $\lambda x^i$
(only the eigenvalue changes, not the eigenspace), it is enough to
restrict attention to the unit sphere $S^2=\{x^i\in {\rm I\!\rm R}^3|\Sigma
x^ix^i=1\}$ in the parameter space. At each point of $S^2$, there is a
two-dimensional, complex vector space ${\rm I \!\!\! \rm C}^2$ of spin states on which
the Hamiltonian acts. Since $H$ squares to unity, its eigenvalues are
$\pm1$. The subspace of positive energy states $\{|v(x)>\in {\cal
C}^2|H(x)|v(x)>=|v(x)>\}$ defines a line bundle over $S^2$ (the Hopf
bundle). To compute the Berry potential which arises when the parameters
$x^i$ are varied, note that $H(x)=h(x)\sigma_3h^{-1}(x)$, where $h(x)$ is
defined by 
\begin{equation} h(x)=\frac{1+H\sigma_3}{\sqrt{2(1+x^3)}},
\label{hofx} \end{equation} 
at all points except the south pole, where
$x^3=-1$. If we pick a normalised positive energy state $|v^N> $ at the 
north pole, which satisfies $\sigma_3|{ v^N}>=|{v^N}>$, $|v(x)>=h(x)|v^N>$ 
is a
normalised positive energy state all over the sphere (except the south
pole).  (Similar considerations also apply to the south polar patch, which
excludes the north pole). The gauge potential describing the connection is
$A=<v(x)d|v(x)>=<v^N|h^{-1}dh|v^N>$ and its field strength is $F=dA$. 
It  is easily seen that this field strength describes a magnetic monopole
situated at the origin of $x^i$, which has been excised from the parameter
space. This magnetic monopole satisfies the Dirac quantisation condition 
with $n=1$. 

Consider the slice
$x^2=0$. In this special case, we can write 
$H(\theta)=\cos\theta\sigma_3+\sin\theta\sigma_1$. Or in the standard
basis for Pauli matrices,

\begin{displaymath}
{\widehat H}(\theta)=
\left(\begin{array}{cc}
\cos\theta&\sin\theta\\
\sin\theta&-\cos\theta
\end{array}\right)
\label{twobytwo}
\end{displaymath}
As $\theta$ goes from $0$ to $2\pi$, the real symmetric matrix $H(\theta)$ 
traverses a loop and returns to itself,
but its eigenvector $(\cos\theta/2,\sin\theta/2)$ reverses sign! This
is just the M\"obius bundle you made with your hands.


\begin{references}
\bibitem{foota} This is why a flat paper label can be stuck 
on the cylindrical surface of a beer bottle without crumpling!
In contrast, try tracing Australia from a globe to a flat piece
of tracing paper.
\bibitem{spivak}
M. A. Spivak, Differential Geometry, Berkeley CA, Publish or Perish inc.
(1999).
\bibitem{milnor}
J. Milnor and J. Stasheff, Characteristic Classes
(Princeton Univerity Press, Princeton, N.J, (1974).
\bibitem{steenrod}
N. Steenrod, Topology of Fibre Bundles (Princeton University
Press, Princeton, N.J. (1951).
\bibitem{kobayashi}
S. Kobayashi and K. Nomizu, Foundations of Differential Geometry
(Interscience, New York, 1963).
\bibitem{wuyang}
T.T. Wu and C.N. Yang, Physical Review {\bf D12}, 3845 (1975).
\bibitem{landau} L.D. Landau and E. M. Lifshitz, ``The Classical Theory of 
Fields'', Pergammon Press, NY, 1986.
\bibitem{geom}
A. Shapere and F. Wilczek, Geometric Phases in Physics,
World Scientific, Singapore (1989).
\bibitem{dna}
Supurna Sinha, `Biopolymer Elasticity', 
Study Circle, November 2003. physics/0308003;
Joseph Samuel and Supurna Sinha, Molecular Elasticity and the 
Geometric Phase, Physical Review Letters, {\bf 90} 098305 (2003).
\bibitem{geomresource}
 Jeeva Anandan, Joy Christian and Kazimir Wanelik,
Resource letter: Geometric phases in physics
Am.J.Phys.{\bf 65}, 180 (1997).
\bibitem{wigner}
E.P. Wigner, ``The unreasonable effectiveness of mathematics in the
natural sciences'', {\it Communications in pure and applied mathematics}
{\bf 13}, No. 1, Feb 1960, N.Y. John Wiley and Sons. 
\end{references}
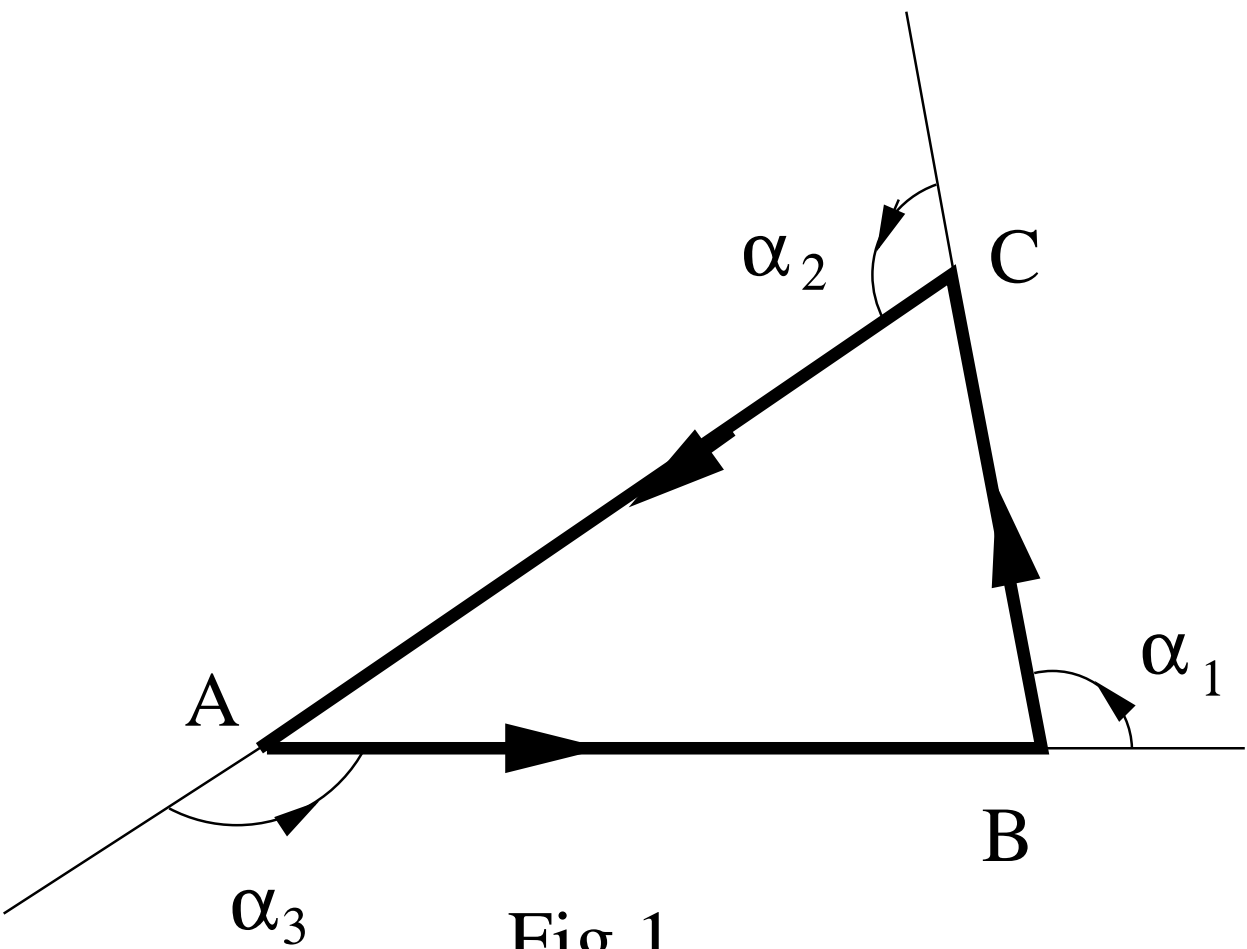
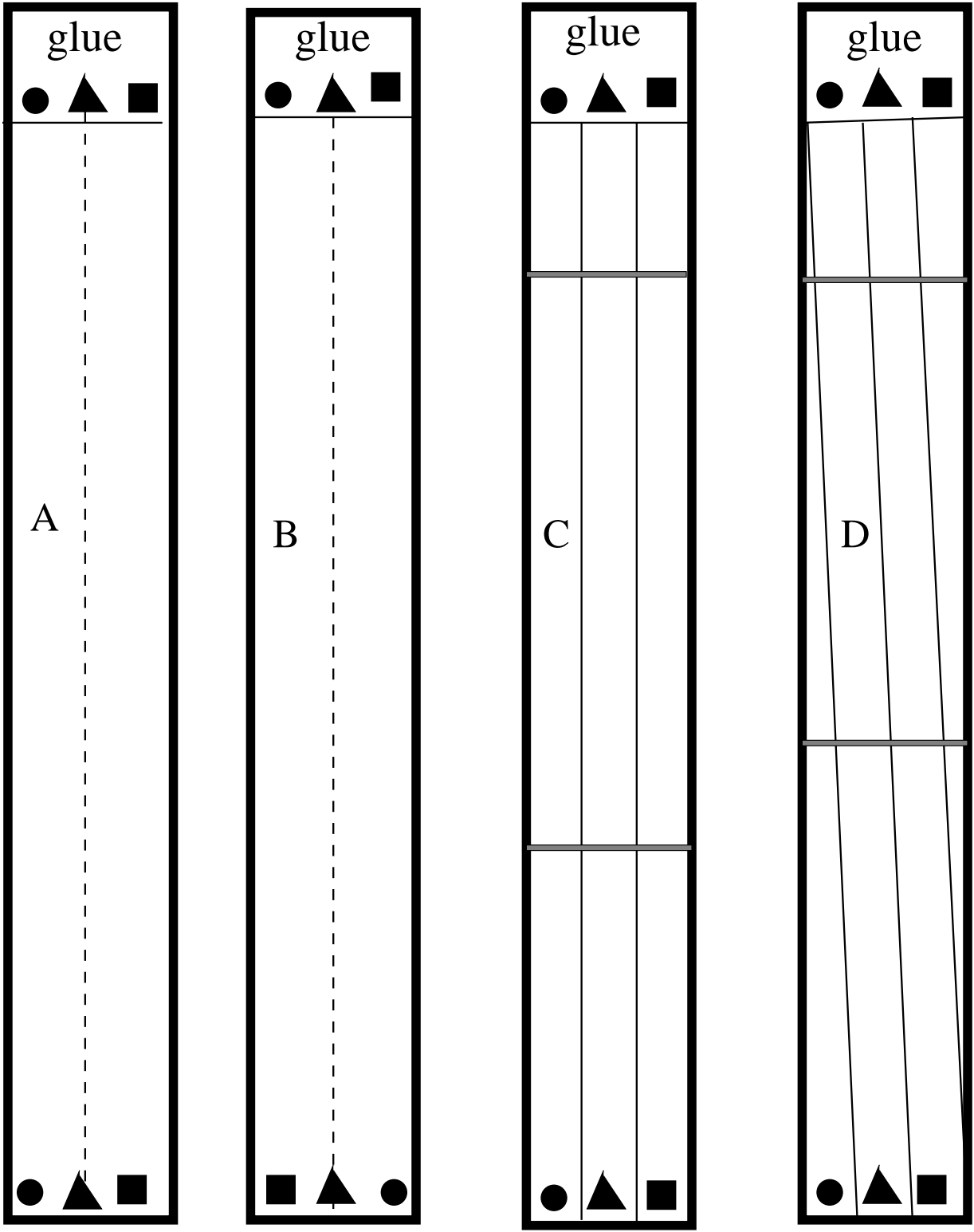
\end{document}